# Untying the insulating and superconducting orders in magic-angle graphene

Petr Stepanov[1], Ipsita Das[1], Xiaobo Lu[1], Ali Fahimniya[2], Kenji Watanabe[3], Takashi Taniguchi[3], Frank H. L. Koppens[1], Johannes Lischner[4], Leonid Levitov[2] and Dmitri K. Efetov[1]*

1. ICFO - Institut de Ciencies Fotoniques, The Barcelona Institute of Science and Technology, Castelldefels, Barcelona, 08860, Spain
2. Department of Physics, Massachusetts Institute of Technology, Cambridge MA 02139, USA
3. National Institute for Materials Science, 1-1 Namiki, Tsukuba, 305-0044, Japan
4. Departments of Materials and Physics and the Thomas Young Centre for Theory and Simulation of Materials, Imperial College London, London SW7 2AZ, United Kingdom

*E-mail: dmitri.efetov@icfo.eu

**The coexistence of superconducting and correlated insulating states in magic-angle twisted bilayer graphene (*1–11*) prompts fascinating questions about the relationship of these orders. Independent control of the microscopic mechanisms governing these phases could help uncover their individual roles and shed light on their intricate interplay. Here we report on direct tuning of electronic interactions in this system by changing its separation from a metallic screening layer (*12,13*). We observe quenching of correlated insulators in devices with screening layer separations that are smaller than a typical Wannier orbital size of 15nm, and with the twist angles slightly deviating from the magic value 1.10° ± 0.05°. Upon extinction of the insulating orders, the vacated phase space is taken over by superconducting domes that feature critical temperatures comparable to those in the devices with strong insulators. In addition, we find that insulators at half-filling can reappear in small out-of-plane magnetic fields of 0.4 T, giving rise to quantized Hall states with a Chern number of 2. Our study suggests reexamination of the often-assumed "mother-child relation" between the insulating and superconducting phases in moiré graphene, and illustrates a new approach to directly probe microscopic mechanisms of superconductivity in strongly-correlated systems.**

Strongly correlated electron systems exhibit a variety of interactions and emergent orders. Famously, the concurrence of unconventional superconductivity (SC) and correlated insulating (CI) phases in cuprates, pnictides and heavy fermion compounds(*14*), has led to the conjecture that superconductivity could be assisted by CI order, therefore arising from a purely electronic mechanism. Achieving a direct control of electron-electron (e-e) interactions – a long-standing goal in the study of correlated electron systems – would clarify the separate origin and the complex relation between these phases. However, previous attempts to control e-e interactions in other crystalline correlated systems were impeded by small atomic orbital sizes and strong sensitivity to doping(*15*).

Magic-angle twisted bilayer graphene (MAG), consisting of two vertically stacked graphene sheets which are slightly misaligned by the magic angle $\theta = \theta_m \sim 1.1°$, has emerged recently as a new correlated system. Flat moiré bands in MAG (*1-11*) and similar "twistronic" systems(*16, 17*) host strongly-correlated electrons that exhibit a variety of interesting ordered states, notably correlated insulators (*1,4,5*), superconductors (*2,3-5*), magnets and topological states (*5-7*). Initial experiments unveiled CIs at integer occupancies of the moiré bands coexisting with superconducting domes appearing upon slight additional doping (*2, 18-22*). The concurrence of these orders was taken as an indication that they are directly related and arise from a common mechanism, similar to the scenarios that have been proposed for the cuprates (*14*).

The exceptionally large moiré unit cell in MAG enables new methods for testing this hypothesis. In the Mott-Hubbard picture, the condition for the appearance of CIs is a large ratio of the on-site Coulomb energy $U$ and the kinetic energy $t$, $U/t \gg 1$. In MAG $t$ can be increased by tuning $\theta$ away from $\theta_m$, which "unflattens" the flat bands. The energy $U$ can be controlled independently by changing the dielectric environment. If the distance $w$ between MAG and a metallic layer is made smaller than the moiré unit cell size, $w < \lambda \sim 15$nm, polarization charges will screen out the Coulomb interactions on that scale and suppress $U$ (*12,13*) (Fig. 1A-B and SI).

Here we report on transport measurements in screening-controlled MAG devices with several near-magic twist angles. We find a strong suppression of the CIs when metallic graphite screening layers are placed closer than 10nm from the MAG plane, separated from it by insulating multilayers of hexagonal boron nitride (hBN), and $\theta$ tuned slightly away from 1.1° by ± 0.05°. Rather than being weakened, SC persists in the absence of the CIs, taking over the phase space vacated by the CIs and spanning across wide doping regions without interruption. This is in stark contrast to all previous studies where metallic layers were placed > 30nm away from the MAG (*1-5*), and SC regions were only found in the presence of CIs. These observations suggest that the insulating and superconducting orders - rather than sharing a common origin - compete with each other, which calls into question a simple analogy with the cuprates.

Our main findings are illustrated in the phase diagrams of 3 representative devices with parameters (D1) $w \sim 7$nm, $\theta \sim 1.15°$; (D2) $w \sim 9.8$nm, $\theta \sim 1.04°$ and (D3) $w \sim 12.5$nm, $\theta \sim 1.10°$; the data are presented in Fig. 1C as color maps and in Fig. 2A as the corresponding line cuts of $\rho_{xx}$ as a function of $v$ and temperature $T$. Fig. 1D shows an atomic force microscope image of a typical device (D1), a quadruple van der Waals stack of graphite/hBN/MAG/hBN, where the bottom hBN layer acts as an ultra-thin $w < 15$nm spacer between the metallic graphite and the MAG. We use four-terminal resistivity $\rho_{xx}$ measurements; the gate voltage $V_g$ on the graphite layer allows us to capacitively tune the carrier density $n$ which is normalized to $n_s$, the density of the fully filled band, which defines the band filling factor $v = n/n_s$. The positions of $v = 0$ mark the charge neutrality point (CNP) of the flat-bands, which in all devices shows a small energy gap. It remains however unclear whether interaction or trivial band effects break the symmetry at the CNP (Extended Data Fig. 6). Superconductivity in all devices was confirmed by a multitude of tests, including measurements of zero resistance and Fraunhofer interference patterns (Fig. 1 E, and ED Fig. 4). High spatial uniformity of our devices (Fig. 1 F, and ED Fig. 3), allows us to rule out that disorder-induced "cloaking" could be responsible for the disappearance of the CIs.

Notably, device D1, which has the thinnest $w$ and only slightly-higher-than-magic $\theta$, demonstrates a phase diagram that is drastically different from all previously reported MAG devices (*2,3-5*). Strikingly, it entirely lacks CI states, showing clearly metallic $\rho_{xx}(T)$ behaviors at $v = \pm 2$, and with $\rho_{xx}$ never exceeding a few k$\Omega$ (Fig. 2A and B). These observations are in line with the absence of Fermi surface reconstruction at $v = \pm 2$, which is evident from the lack of a new sequence of quantum oscillations originating from this point, as well as the very weak Hall density offset in perpendicular magnetic field (ED Fig. 5 and 7 ). Nonetheless, we observe two broad SC domes in both the valence and conduction moiré bands, with values of $T_c \approx 920$ mK and $T_c \approx 420$ mK, respectively (Fig. 1E). While these SC states resemble those reported previously, they are not accompanied by CIs and span across the integer filling positions.

This behavior is in stark contrast to D3, which has the thickest $w$ and a $\theta$ which corresponds exactly to $\theta_m$. It shows resistance peaks at all integer CI fillings and strong thermally activated transport (Fig. 2A and B), which are accompanied by a new set of quantum oscillations and Hall density offsets in magnetic field. Here the SC domes directly flank the CIs with $T_c$ values that range from 150mK to 3K, however, at integer fillings CIs always persist. Device D2, with an intermediate $w$ value and a $\theta$ slightly smaller than $\theta_m$, displays features present in both D1 and D3. While it does not display CIs in the valence band, it shows a single SC dome with a $T_c \approx 400$ mK. In the conduction band it features a not activated resistance peak at $v = 2$, suggesting an underdeveloped CI, and two SC domes flanking it, with $T_c \approx 500$ mK and $T_c \approx 650$ mK.

Overall these findings clearly show that SC states in MAG can exist independently of CIs, indicating that these phases are in competition with one another rather than share a common microscopic origin. While insulating states are quenched in D1 and D2, the corresponding superconducting $T_c$ values remain virtually unaffected, falling in the same $T_c \sim 500$mK – 1.5K range which was previously reported for the devices with similar twist-angles, but in the presence of strong insulators(*2, 3-5*). Since both the screening and band-width effects affect $U/t$, it is difficult to completely disentangle these effects from the data sets presently available. Comparison with the limited data from the literature (*1–5*), however, indicates that screening may be a dominant effect in D1 ($\theta \sim 1.15°$) and D2 ($1.04°$), as these are the only reported devices (*1,3-5*) which do not show any signatures of CI states so close to the magic angle of 1.1°.

We further examine the transport properties of device D1 in a perpendicular magnetic field $B$. For $B > 0.3$T an insulating gap $\Delta$ develops at exactly $v = -2$ filling, which is evident from a transition from a metallic to an activated transport behavior (Fig. 3A). Fitting $R_{xx} \sim exp(\Delta/2 k_b T)$ as a function of $B$, shows an approximately linear evolution of $\Delta$ (here $k_b$ is the Boltzmann constant). The observed magnetoresistance can be attributed to field-induced quenching of the electron dynamics, where the magnetic field "freezes" the kinetic energy $t$ of the electrons by creating Aharonov-Bohm flux through each moiré super-cell (*23-27*). This can tip the balance between the kinetic energy and Coulomb interaction in favor of the latter, producing an increase of the $U/t$ ratio and the recurrence of the CI order.

Strikingly, at the same low $B$-fields at which the CI appears, a quantized Hall state develops. At fields as low as $B \sim 0.4$T we observe regions of near-perfectly quantized plateaus of Hall resistance $R_{xy} \sim h/2e^2 = 12.9$ k$\Omega$ (with $e$ the electron charge and $h$ Planck's constant), accompanied by a vanishing longitudinal resistance $R_{xx} \sim 0$ $\Omega$ (Fig. 3B and C). These values are consistent with the degeneracy of the expected CI at half-filling (*1*) and are two times smaller than the free-electron values. Starting from $v = -2$ at $B = 0$, this state forms in a broad wedge-like region in the $n$-$B$ phase space, extending into larger densities as $B$ increases; the wedge boundary follows a $dn/dB = 2e/h$ dependence which is consistent with a Chern number of 2 (Fig. 3D and E).

The $B$-field at which this quantized Hall state appears is almost an order of magnitude lower than the one needed to quantize all other quantum Hall plateaus in the system, $B > 3$T (ED Fig. 7). This state appears in the absence of single-particle Landau level quantization in the system, which suggests a different mechanism, probably originating from e-e interactions. This scenario is supported by the appearance of the symmetry broken CI state at $v = -2$ at the same $B$-field. Theory predicts that strong correlations can favor a valley-polarized Chern insulator even for untwisted bilayer graphene(*28,29*), however in topological flat-bands in MAG correlations are expected to boost and further strengthen the Chern insulator order (*30*). Similar observations were recently made in hBN-aligned(*6,7*) and non-aligned MAG(*5*) and in ABC trilayer graphene(*16*) for odd-integer filled states, where orbital magnetic states were proposed. Sizeable hysteresis around $B \sim 0.4$T in device D1 potentially signals the existence of a similar state (*19*).

What are the implications of these findings for the origin of the superconducting state? The observed resilience of superconductivity upon suppression of the insulating phase is consistent with the two phases competing rather than being intimately connected. Such competition would be hard to reconcile with a common microscopic mechanism of the two phases suggested by an analogy with cuprates. Instead, it appears that Coulomb interactions drive the formation of the commensurate insulators, whereas superconductivity arises from a more conventional mechanism. However, the anomalous character of superconductivity in MAG, occurring at record-low carrier densities, suggests that the electron-phonon mechanism, if present, is enhanced by the high density of states and electron correlation effects. The reappearance of the correlated insulator phases in abnormally weak magnetic fields confirms that correlations remain strong in the system, calling for a better understanding of their impact on the superconducting order.

**References**

1. Y. Cao, V. Fatemi, A. Demir, S. Fang, S. L. Tomarken, J. Y. Luo, J. D. Sanchez-Yamagishi, K. Watanabe, T. Taniguchi, E. Kaxiras, R. C. Ashoori, P. Jarillo-Herrero, Correlated insulator behaviour at half-filling in magic-angle graphene superlattices. *Nature*. **556**, 80–84 (2018).
2. Y. Cao, V. Fatemi, S. Fang, K. Watanabe, T. Taniguchi, E. Kaxiras, P. Jarillo-Herrero, Unconventional superconductivity in magic-angle graphene superlattices. *Nature*. **556**, 43–50 (2018).
3. K. Kim, A. DaSilva, S. Huang, B. Fallahazad, S. Larentis, T. Taniguchi, K. Watanabe, B. J. LeRoy, A. H. MacDonald, E. Tutuc, Tunable moiré bands and strong correlations in small-twist-angle bilayer graphene. *Proc. Natl. Acad. Sci.* **114**, 3364–3369 (2017).
4. M. Yankowitz, S. Chen, H. Polshyn, Y. Zhang, K. Watanabe, T. Taniguchi, D. Graf, A. F. Young, C. R. Dean, Tuning superconductivity in twisted bilayer graphene. *Science* **363**, 1059–1064 (2019).
5. X. Lu, P. Stepanov, W. Yang, M. Xie, M. A. Aamir, I. Das, C. Urgell, K. Watanabe, T. Taniguchi, G. Zhang, A. Bachtold, A. H. MacDonald, D. K. Efetov, Superconductors, Orbital Magnets, and Correlated States in Magic Angle Bilayer Graphene. *Nature*. **574**, 653–657 (2019).
6. A. L. Sharpe, E. J. Fox, A. W. Barnard, J. Finney, K. Watanabe, T. Taniguchi, M. A. Kastner, D. Goldhaber-Gordon, Emergent ferromagnetism near three-quarters filling in twisted bilayer graphene, *Science* **365**, 605-608 (2019).
7. M. Serlin, C. L. Tschirhart, H. Polshyn, Y. Zhang, J. Zhu, K. Watanabe, T. Taniguchi, L. Balents, A. F. Young, Intrinsic quantized anomalous Hall effect in a moiré heterostructure, arXiv:1907.00261 (2019).
8. Yonglong Xie, Biao Lian, Berthold Jäck, Xiaomeng Liu, Cheng-Li Chiu, Kenji Watanabe, Takashi Taniguchi, B. Andrei Bernevig & Ali Yazdani, Spectroscopic signatures of many-body correlations in magic-angle twisted bilayer graphene, *Nature* **572**, 101 (2019).
9. Alexander Kerelsky, Leo J. McGilly, Dante M. Kennes, Lede Xian, Matthew Yankowitz, Shaowen Chen, K. Watanabe, T. Taniguchi, James Hone, Cory Dean, Angel Rubio & Abhay N. Pasupathy, Maximized electron interactions at the magic angle in twisted bilayer graphene, *Nature* **572**, 95–100 (2019).
10. Yuhang Jiang, Xinyuan Lai, Kenji Watanabe, Takashi Taniguchi, Kristjan Haule, Jinhai Mao & Eva Y. Andrei, Charge order and broken rotational symmetry in magic-angle twisted bilayer graphene, *Nature* **573**, 91–95 (2019).

11. Youngjoon Choi, Jeannette Kemmer, Yang Peng, Alex Thomson, Harpreet Arora, Robert Polski, Yiran Zhang, Hechen Ren, Jason Alicea, Gil Refael, Felix von Oppen, Kenji Watanabe, Takashi Taniguchi & Stevan Nadj-Perge, Electronic correlations in twisted bilayer graphene near the magic angle, *Nature Physics* **15**, 1174–1180 (2019).
12. Z. A. H. Goodwin, F. Corsetti, A. A. Mostofi, J. Lischner, Twist-angle dependence of electron correlations in moiré graphene bilayers. *Phys. Rev. B*. **100**, 121106 (2019).
13. J. M. Pizarro, M. Rösner, R. Thomale, R. Valentí, T. O. Wehling, Internal screening and dielectric engineering in magic-angle twisted bilayer graphene. *Phys. Rev. B*. **100,** 161102 (2019).
14. Patrick A. Lee, Naoto Nagaosa, and Xiao-Gang Wen, Doping a Mott insulator: Physics of high-temperature superconductivity, *Rev. Mod. Phys.* **78**, 17 (2006).
15. F. Nilsson, K. Karlsson, and F. Aryasetiawan, Dynamically screened Coulomb interaction in the parent compounds of hole-doped cuprates: Trends and exceptions, *Phys. Rev. B* **99**, 075135 (2019).
16. Guorui Chen, Aaron L. Sharpe, Eli J. Fox, Ya-Hui Zhang, Shaoxin Wang, Lili Jiang, Bosai Lyu, Hongyuan Li, Kenji Watanabe, Takashi Taniguchi, Zhiwen Shi, T. Senthil, David Goldhaber-Gordon, Yuanbo Zhang, Feng Wang, Tunable Correlated Chern Insulator and Ferromagnetism in Trilayer Graphene/Boron Nitride Moiré Superlattice, *arXiv:1905.06535* (2019).
17. X. Liu, Z. Hao, E. Khalaf, J. Y. Lee, K. Watanabe, T. Taniguchi, A. Vishwanath, P. Kim, Spin-polarized Correlated Insulator and Superconductor in Twisted Double Bilayer Graphene (2019) (available at http://arxiv.org/abs/1903.08130).
18. M. Ochi, M. Koshino, K. Kuroki, Possible correlated insulating states in magic-angle twisted bilayer graphene under strongly competing interactions, *Phys. Rev. B*. **98**, 081102 (2018).
19. M. Xie, A. H. MacDonald, On the nature of the correlated insulator states in twisted bilayer graphene, *arXiv:1812.04213* (2018).
20. J. F. Dodaro, S. A. Kivelson, Y. Schattner, X. Q. Sun, C. Wang, Phases of a phenomenological model of twisted bilayer graphene. *Phys. Rev. B*. **98,** 075154 (2018).
21. H. C. Po, L. Zou, T. Senthil, A. Vishwanath, Faithful tight-binding models and fragile topology of magic-angle bilayer graphene. *Phys. Rev. B*. **99**, 195455 (2019).
22. L. Zou, H. C. Po, A. Vishwanath, T. Senthil, Band structure of twisted bilayer graphene: Emergent symmetries, commensurate approximants, and Wannier obstructions. *Phys. Rev. B*. **98**, 085435 (2018).
23. C. R. Dean, L. Wang, P. Maher, C. Forsythe, F. Ghahari, Y. Gao, J. Katoch, M. Ishigami, P. Moon, M. Koshino, T. Taniguchi, K. Watanabe, K. L. Shepard, J. Hone & P. Kim, Hofstadter's butterfly and the fractal quantum Hall effect in moiré superlattices, *Nature* **497**, 598–602 (2013).
24. B. Hunt, J. D. Sanchez-Yamagishi,, A. F. Young, M. Yankowitz, B. J. LeRoy, K. Watanabe, T. Taniguchi, P. Moon, M. Koshino, P. Jarillo-Herrero, R. C. Ashoori, Massive Dirac Fermions and Hofstadter Butterfly in a van der Waals Heterostructure, *Science* **340**, 6139 (2013).
25. L. A. Ponomarenko, R. V. Gorbachev, G. L. Yu, D. C. Elias, R. Jalil, A. A. Patel, A. Mishchenko, A. S. Mayorov, C. R. Woods, J. R. Wallbank, M. Mucha-Kruczynski, B. A. Piot, M. Potemski, I. V. Grigorieva, K. S. Novoselov, F. Guinea, V. I. Fal'ko & A. K. Geim, Cloning of Dirac fermions in graphene superlattices, *Nature* **497**, 594–597 (2013).
26. R. Nandkishore, L. Levitov, Dynamical screening and excitonic instability in bilayer graphene. *Phys. Rev. Lett.* **104**, 156803 (2010).
27. R. Nandkishore, L. Levitov, Quantum anomalous Hall state in bilayer graphene. *Phys. Rev. B* **82**, 115124 (2010).
28. R. T. Weitz, M. T. Allen, B. E. Feldman, J. Martin, A. Yacoby, Broken-Symmetry States in Doubly Gated Suspended Bilayer Graphene, *Science* **330**, 812 (2010).
29. A. F. Young, J. D. Sanchez-Yamagishi, B. Hunt, S. H. Choi, K. Watanabe, T. Taniguchi, R. C. Ashoori and P. Jarillo-Herrero, Tunable symmetry breaking and helical edge transport in a graphene quantum spin Hall state, *Nature* **505**,528–532 (2014).
30. N. Bultinck, E. Khalaf, S. Liu, S. Chatterjee, A. Vishwanath and M. P. Zaletel, Ground State and Hidden Symmetry of Magic Angle Graphene at Even Integer Filling, *arXiv:1911.02045* (2019).
31. K. Kim, A. DaSilva, S. Huang, B. Fallahazad, S. Larentis, T. Taniguchi, K. Watanabe, B. J.

LeRoy, A. H. MacDonald, E. Tutuc, Tunable moiré bands and strong correlations in small-twist-angle bilayer graphene. Proc. Natl. Acad. Sci. 114, 3364–3369 (2017).

Acknowledgements:
We are grateful for fruitful discussions with Pablo Jarillo-Herrero, Allan MacDonald, Andrea Young, Leon Balents, Andrei Bernevig, Biao Lian, Cory Dean. D.K.E. acknowledges support from the Ministry of Economy and Competitiveness of Spain through the "Severo Ochoa" program for Centres of Excellence in R&D (SE5-0522), Fundació Privada Cellex, Fundació Privada Mir-Puig, the Generalitat de Catalunya through the CERCA program, the H2020 Programme under grant agreement n° 820378, Project: 2D·SIPC and the La Caixa Foundation. L.L. acknowledges support from the Science and Technology Center for Integrated Quantum Materials, NSF Grant No. DMR-1231319; and Army Research Office Grant W911NF-18-1-0116. F.H.L.K. acknowledges support from the ERC consolidator grant TOPONANOP (grant agreement n° 726001). K.W. and T.T. acknowledge support from the Elemental Strategy Initiative conducted by the MEXT, Japan and and the CREST (JPMJCR15F3), JST.

Author contributions:
D.K.E, X.L. and P.S. conceived and designed the experiments; P.S., X.L. and I.D. performed the experiments; P.S. and D.K.E. analyzed the data; A.F. and L.L. performed the theoretical modeling; T.T. and K.W. contributed materials; D.K.E. and F. H.L.K. supported the experiments: D. K.E, L.L. and P.S. wrote the paper.

**Supplementary Information** is available for this paper.

**Correspondence and requests for materials** should be addressed to D.K.E.

**Reprints and permissions information** is available at *www.nature.com/reprints*.

**Competing financial and non-Financial interests**:
The authors declare no competing financial and non-financial interests.

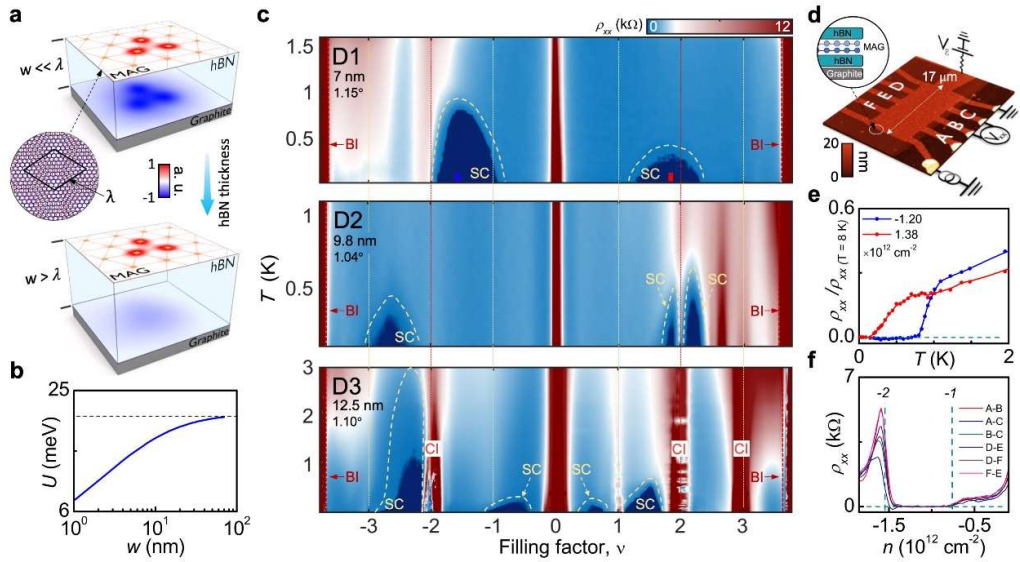

**Fig. 1. | Screening-controlled MAG phase diagrams for near-magic twist angles. a**, Wannier orbitals (red) in MAG are screened by image charges on the graphite surface (blue). **b**, Calculated on-site Coulomb energy $U$ vs. hBN spacer thickness $w$ for $\theta = 1.05°$ (dashed line marks the unscreened value). **c**, Color plot of resistivity $\rho_{xx}$ vs. moiré band filling factor $v$ and temperature $T$ for three screening-controlled MAG devices (D1) $w \sim 7$nm, $\theta \sim 1.15°$; (D2) $w \sim 9.8$nm, $\theta \sim 1.04°$ and (D3) $w \sim 12.5$nm, $\theta \sim 1.10°$. Signatures of CIs are completely absent in devices with the thinnest $w$ (D1 and D2), while SC persists and $T_c$ values remain virtually unaffected. "BI" denotes band insulator. **d**, AFM image and measurement scheme for device D1, a graphite/hBN/MAG/hBN stack in which the graphite flake acts as the screening layer as well as the back gate. **e**, Normalized $\rho_{xx}$ vs. $T$ shows SC transitions for both SC domes in D1, taken at carrier densities $n$ that are marked in **c**. **f**, Four-probe resistivity $\rho_{xx}$ vs. $n$ for different contact pairs of D1 shows homogeneous distribution of SC across the entire device.

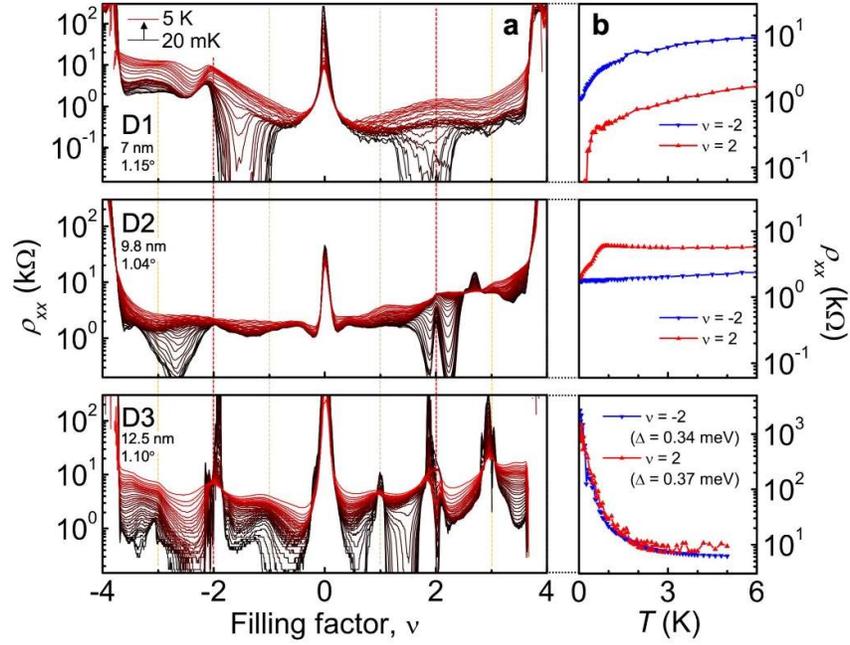

**Fig. 2. | The SC and CI phases temperature and density dependence. a**, Line cuts of resistivity $\rho_{xx}$ vs. filling factor $\nu$, of devices D1-D3, for different temperatures $T$ from 20 mK to 5 K. **b**, $\rho_{xx}$ vs. $T$ for moiré band fillings of $\nu = \pm 2$ for each device. While D1 and D2 show metallic behavior in both valence and conduction bands, D3 shows strongly thermally activated behavior $\rho_{xx} \sim exp(\Delta/2\, k_b T)$ consistent with strong CI order.

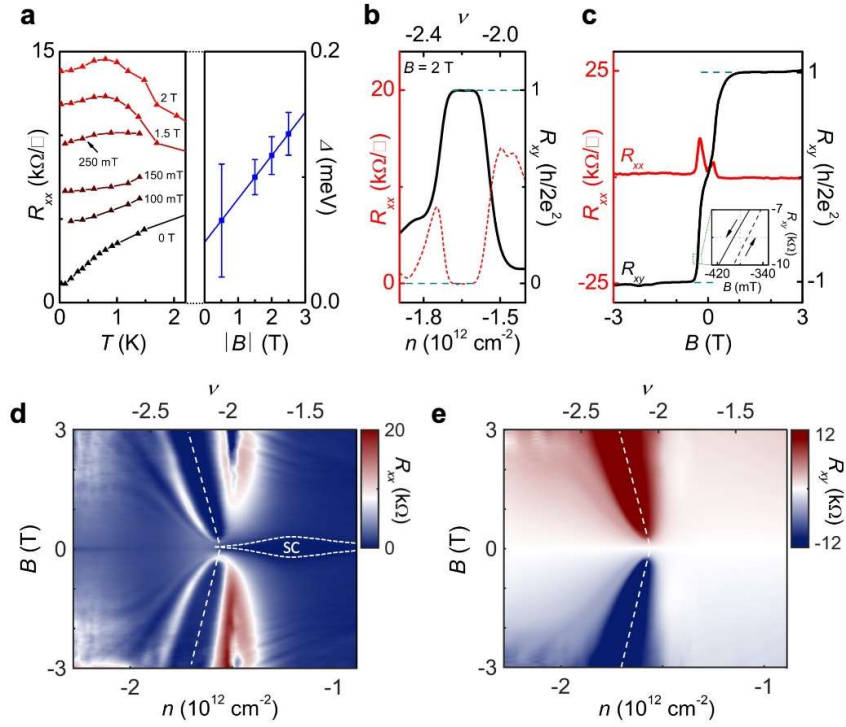

**Fig. 3. | Formation of Chern insulators in small out of plane magnetic field. a**, Left panel shows longitudinal sheet resistance $R_{xx}$ vs. temperature $T$ for moiré band filling $\nu = -2$ for $B$-fields from 0T to 2T. It reveals a transition from a metallic to a thermally activated behavior $R_{xx} \sim exp(\Delta/2\,k_bT)$, where the energy gap $\Delta$ increases linearly with $B$ (right panel). **b** and **c**, $R_{xx}$ and transverse resistance $R_{xy}$ vs. charge carrier density $n$ at $B = 2$T (**b**) and vs. $B$ at $n = -1.75\times10^{12}$ cm$^{-2}$ (**c**), shows quantization of $R_{xy} = $ h/2e$^2$ and $R_{xx} = 0\,\Omega$ above $|B| \geq 0.4$ T. **d** and **e**, Color plots of $R_{xx}$ (**d**) and $R_{xy}$ (**e**) vs. $n$ and $B$ demonstrate the developing wedge-shaped phase diagram of Chern insulator.

**Methods**:

Screening layer fabrication process.
The samples are fabricated using the "cut-and-stack" method, in analogy to the previously introduced "tear-and-stack" technique (*31*). Typically, a thin hBN flake is picked up by a propylene carbonate (PC) film, which is placed on a polydimethyl siloxane (PDMS) stamp at 90 ºC (ED Fig. 1). The hBN flake is then used to pick up a part of a pre-cut monolayer graphene flake, mechanically exfoliated on $Si^{++}/SiO_2$ (285 nm) surface. Subsequently, the second half of a graphene sheet is rotated to a target angle usually around $\theta \sim 1.1º$ - $1.15º$ and then picked up by hBN/graphene stack on PC from the previous step. The heterostructure is then placed on top of another thin hBN flake. Usually, the bottom hBN flake thickness is chosen by optical contrast and further confirmed with atomic force microscopy (AFM) measurements. The very bottom layer of the heterostructure consists of a relatively thick graphite flake (typically a few layer graphene > 1nm thick) that acts as a local back gate and a screening layer simultaneously. The final stack is then placed on a target $Si^{++}/SiO_2$ (285 nm) wafer, where it is further etched into a multiple Hall bar geometry using $CHF_3/O_2$ plasma and edge-coupled to Cr/Au (5/50 nm) metal contacts.

Dielectric thickness measurements.
Bottom hBN thickness is obtained from atomic force microscopy (AFM) measurements. ED Fig. 2 demonstrates a set of MAG heterostructures that have been used to fabricate the devices reported in the main text. The upper panel shows an optical image of the final graphite /hBN/MAG/hBN stack. We find that the heterostructures exhibit high structural homogeneity and do not show visible bubble formations, which are known to locally distort the twist-angle and charge carrier density. These observations are further confirmed by the AFM scans shown in the insets. The AFM scans are also used to extract the topography of the fabricated stacks where we find hBN thicknesses of $w \sim 7.0$ nm (D1), 9.8 nm (D2) and 12.5 nm (D3).

The difference in bottom dielectric thickness is further confirmed by measurements of the capacitance between the graphite and the MAG layers. Extracted from quantum oscillations map (ED Fig. 7), we find that the back gate capacitance changes from 355 $nF/cm^2$ (D1) to 260 $nF/cm^2$ (D2) and further to 221 $nF/cm^2$ (D3), in a good agreement with the extracted AFM height profiles.

Effects of disorder in our devices.
During the stacking procedure, twist angle disorder may be introduced into the final heterostructure. To further analyze the twist angle homogeneity, we perform two-terminal conductance measurements vs. carrier density for different sets of contact pairs, as shown in ED Fig. 3 for device D1. Contact pairs on the right-hand side of the sample (A-B, B-C, C-D, and D-E) demonstrate the highest angle homogeneity with maximum charge carrier density deviation of the full filled superlattice unit cell $\delta n_s^{max} \sim 0.03 \times 10^{12}$ $cm^{-2}$. Among them, contact pair C-D has been used to obtain longitudinal resistance data $R_{xx}$ in this device. Overall, the graph demonstrates high global twist angle homogeneity as the positions of the band-insulator gaps match exceptionally well. The position of the charge neutrality point (CNP) varies little between the different contact pairs ($V_g^{CNP} = -4 \pm 1$ mV).

In addition, we perform the analysis on the full-width-half-maximum (FWHM) of the resistance peak at the CNP, and the sharpness of the band-gap edges at complete filling of the moiré bands. We estimate a charge carrier inhomogeneity at CNP of $\delta n \sim 1.8 \times 10^{10}$ $cm^{-2}$, which suggests that the sample inhomogeneity is mainly defined by charge disorder. Moreover, all devices show very small twist-angle inhomogeneity, with variation as small as $\Delta\theta \sim 0.01°$ per 10μm (SI), with robust, macroscopic superconducting regions spanning the entire device area, which was confirmed by transport measurements using different contact pairs across the device (Fig. 1 F). Here we note, that in a situation where disorder would mix the SC and CI phases in the sample, a percolating SC network could in principle short circuit the CI phases. However, the exceptionally low level of the charge inhomogeneity and twist-angle disorder makes such a scenario highly unlikely.

Superlattice density extraction.
The phase diagram shown in ED Fig. 7 is used to estimate the twist angles in the measured devices. We use the relation $n_s/\nu \sim 2\theta^2/\sqrt{3}a^2$, where $a = 0.246$ nm is the lattice constant of graphene and $n_s$ is the charge carrier density corresponding to a fully filled superlattice unit cell. Quantum oscillations propagating outside of the full filled flat band (e.g. black solid and dashed lines in ED Fig. 7) converge at a point on the horizontal axis that defines $n_s$. Using this position, we find that $n_s \sim 3.04\times10^{12}$ cm$^{-2}$ (or in the absolute gate voltage values $V_g - V_g^{CNP} \approx$ 1.38 V). This constitutes a twist angle in D1 $\theta \sim 1.15°$.

Transport measurements.
All measurements were carried out in a dilution refrigerator in out-of-plane magnetic field $B$. For transport measurements we have employed a standard low frequency lock-in technique using Stanford Research SR860 amplifiers with excitation frequency $f = 19.111$ Hz. For controlling back gate voltage, we use Keithley 2400s voltage source-meters. DC voltage vs. DC excitation current measurements are performed using SR560 low-noise DC voltage preamplifier in combination with a Keithley 2700 multimeter. In order to achieve a lower electron temperature in our measurements, we 1) perform electronic filtering of the measurement setup using a network of commercially available low-pass RC and RF filters and 2) use very low excitation currents (<10 nA) due to a risk of overheating electrons and fragility of SC phases. AC excitation currents are applied through a 10 MΩ resistor. All data in this study is taken at the fridge base temperature ~20-30 mK unless specified.

**Data availability**:

The data that support the findings of this study are available from the corresponding author upon reasonable request.

**Extended data figure legends**:

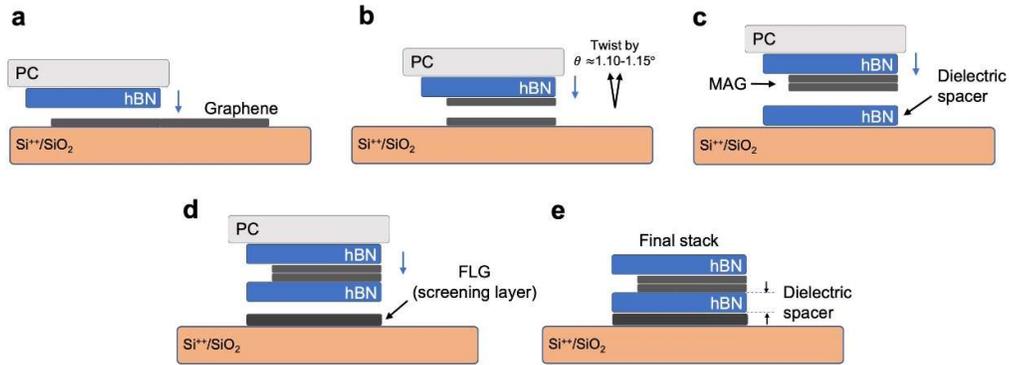

**Extended Data Fig. 1 | Screening layer fabrication method. a**, Sacrificial PC layer is used to pick up top hBN, which is further used to pick up the first half of a monolayer graphene flake. **b**, The second half of the graphene flake is rotated by 1.10º - 1.15º and subsequently picked up by the hBN/graphene stack on PC. **c**, The heterostructure is further placed on the bottom hBN, which is served as a dielectric spacer. **d-e,** At the final step the heterostructure is placed on top of the target wafer and the PC layer is removed in chloroform.

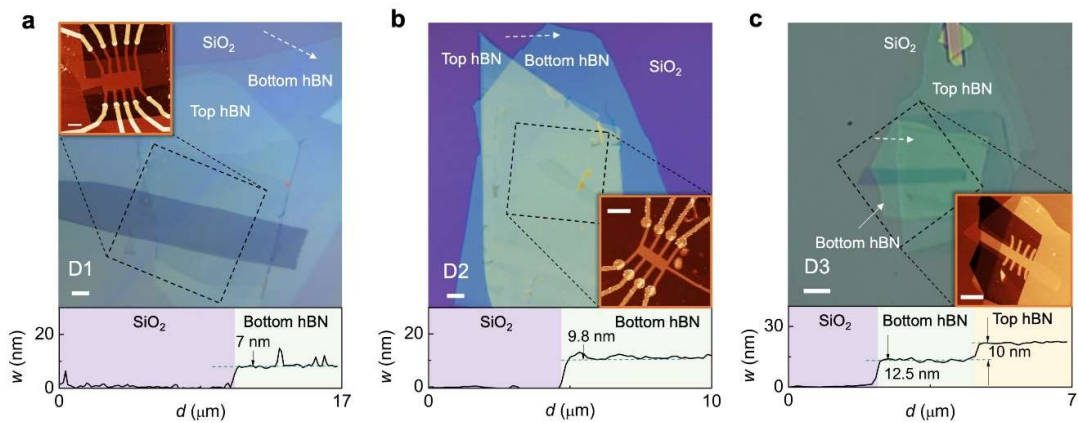

**Extended Data Fig. 2 | AFM and optical micrographs for samples D1(a), D2(b), and D3(c) reported in the main text.** The main panels are optical images of final stacks, from which all three devices were fabricated. The insets demonstrate AFM scans of the final devices etched into multi-terminal Hall bar geometries. Dashed black squares show AFM image areas. Bottom hBN thickness measurements are shown on the lower panel graphs. Height profiles are taken along the white dashed arrow lines. Scale bars are 5 μm.

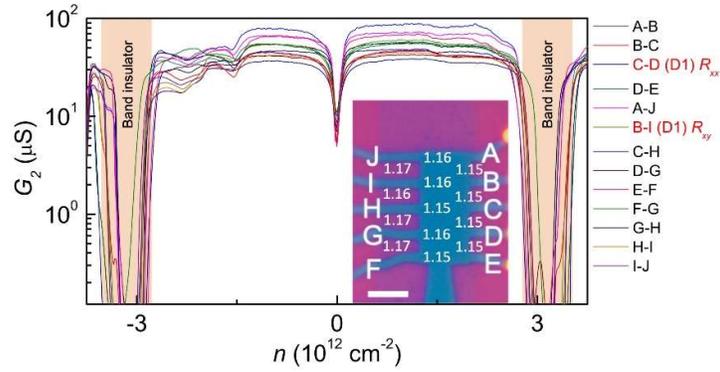

**Extended Data Fig. 3 | Two-terminal conductance measurements across all available contacts in device D1.** The legend shows contact pairs corresponding to the inset optical image of the device. Numbers on the device optical micrograph correspond to measured global twist angle values between contact pairs (extracted from resistance maxima). Scale bar is 5 μm. Data is taken at 20 mK.

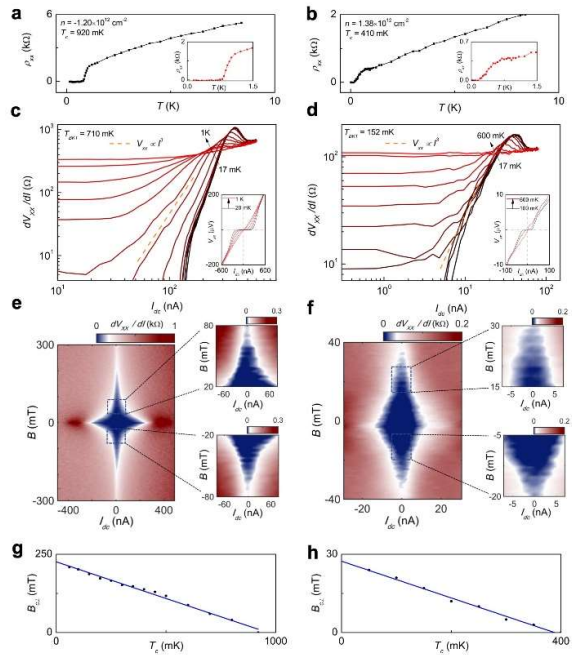

**Extended Data Fig. 4 | SC states characterization in device D1.** The right- (left-) hand side of the figure refers to the SC pocket in the valence (conductance) band on the top panel of Fig. 1c in the main text. **a-b**, Temperature activation of superconductivity for both pockets on the absolute resistivity scale. The insets demonstrate a zoomed-in range of temperatures from 0 K to 1.5 K. **c-d**, Berezinskii–Kosterlitz–Thouless (BKT) measurements of differential resistance $dV_{xx}/dI$ versus DC current bias $I_{dc}$ for both SC pockets. The insets show DC voltage as a function DC current bias taken at different temperatures for optimally doped SC states. **e-f**, Differential conductance $dV_{xx}/dI$ (color) as a function of perpendicular magnetic field $B$ and DC current bias $I_{dc}$ shows distinct diamond-like features for both pockets. Zoomed-in images (to the right) show clear Fraunhofer interference patterns, which are a solid proof of superconductivity. **g-h**, Ginzburg-Landau coherence length measurements for both pockets. Critical field $B_{c\perp}$ versus critical temperature $T_c$ taken at half of normal state resistance values. Black dots refer to experimentally obtained values, blue lines are linear fit to the data. We estimate coherence length $\xi_{GL}$ = 38 nm (**g**) and $\xi_{GL}$ = 101 nm (**h**).

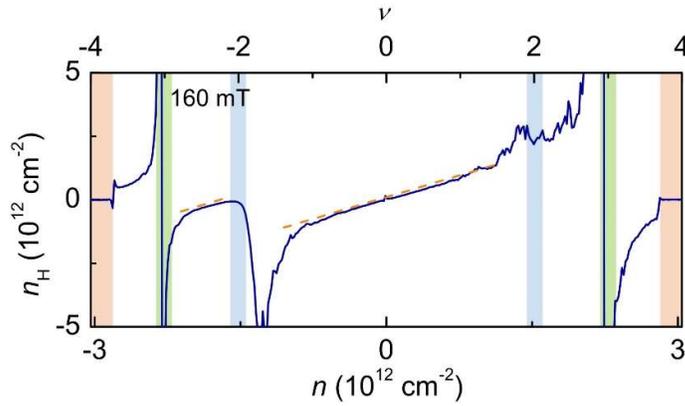

**Extended Data Fig. 5 | Hall density measurements in device D1.** Hall density $n_H$ versus charge carrier density $n$ extracted from the low-field Hall resistance measurements at 160 mT.

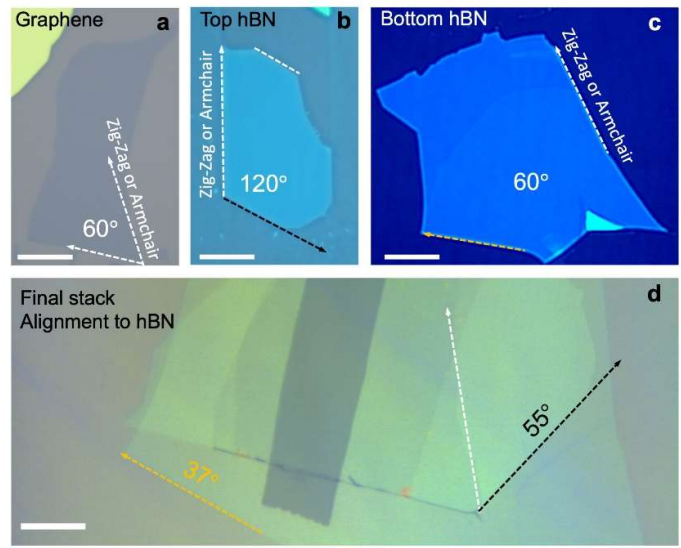

**Extended Data Fig. 6 | Check for alignment to hBN. a**, Optical image of the monolayer graphene flake on $SiO_2$ substrate used for fabrication of MAG heterostructure. White dashed arrows indicate preferable lattice directions (zig-zag or armchair). **b**, Optical image of the top hBN flake. **c**, Optical micrograph of the bottom hBN flake. **d**, Optical image of the final stack on $SiO_2$ substrate. Black dashed arrow indicates the edge of top hBN shown in **b**. Orange dashed arrow corresponds to the edge of the bottom hBN shown in panel **c**. Black (orange) numbers correspond to angle between white dashed arrow (graphene edge) and top (bottom) hBN. We estimate twist angle between bottom hBN and MAG ~ 7 (±1.5)° or ~ 23 (±1.5)° and between top hBN and MAG ~ 25 (±1.5)° or ~ 5 (±1.5)°. Furthermore, we don't find signatures of alignment to hBN in magnetic field SdH oscillation data for neither of three devices (e.g. Extended Data Fig. 7). Scale bars are 20 μm.

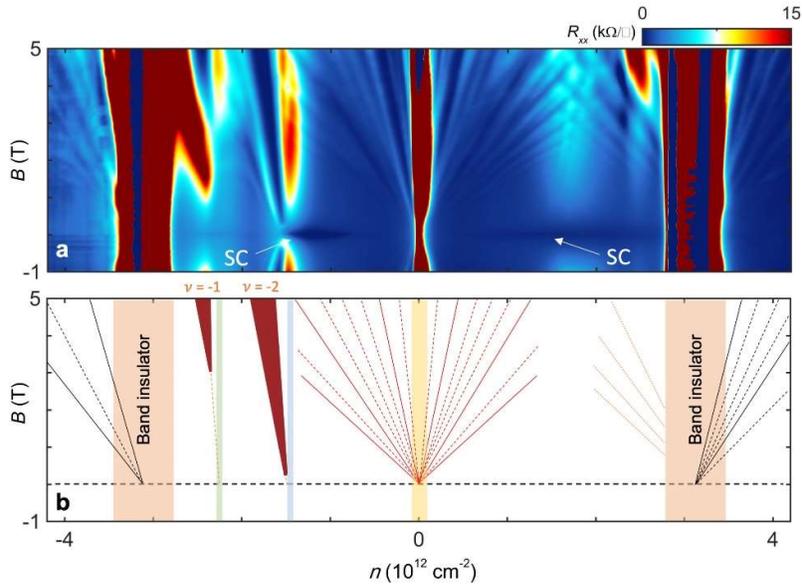

**Extended Data Fig. 7 | Full range magnetic field phase diagram in device D1 at 30 mK. a**, Longitudinal sheet resistance $R_{xx}$ (color) versus change carrier density $n$ and perpendicular magnetic field $B$. **b**, Schematic image of Landau levels shown in **a**. Solid lines correspond to 4-fold degenerate levels with quantized plateaus $\nu_{\square\square}$ = 4, 8, 12, … Dashed lines show broken spin or/and valley degeneracy levels. Dark red features to the left show Chern-like insulator states originating from $\nu$ = -2 (-3)-filled superlattice unit cell corresponding to quantization of $2e^2/h$ ($e^2/h$). Blue and green transparent stripes on the left side of the graph correspond to superlattice unit cell with commensurate fillings $\nu$ = -2 and -3, respectively.

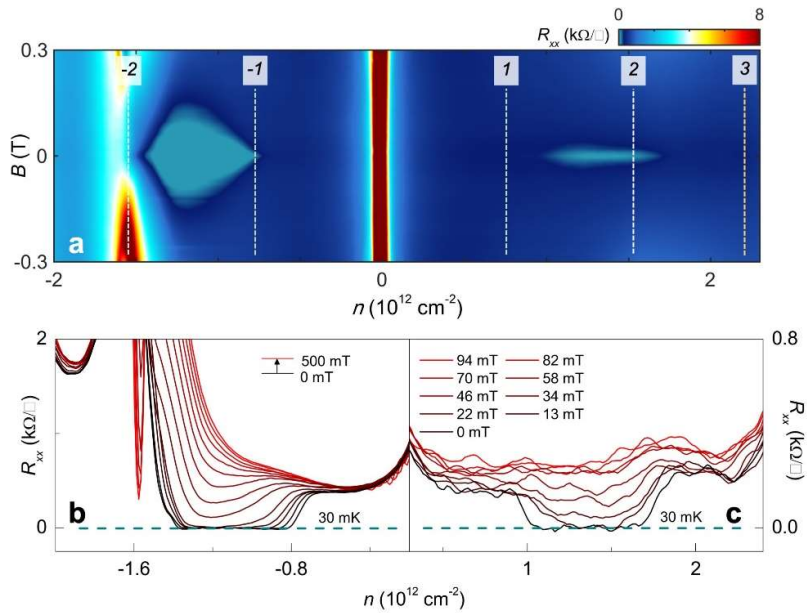

**Extended Data Fig. 8 | Effect of low magnetic field on correlated states in device D1. a**, Longitudinal sheet resistance $R_{xx}$ (color) versus charge carrier density $n$ and perpendicular magnetic field $B$. $I_{ac}$ =10 nA. **b**, $R_{xx}$ as a function of charge carrier density $n$ for valence band SC pockets in **a**. $I_{ac}$ =10 nA. **c**, Resistance versus charge carrier density for valence band SC pockets in **a**. $I_{ac}$ =1 nA.

## Supplementary Information

**Modeling Screening of Coulomb interactions in graphene/hBN/graphite heterostructures.**
Here we discuss the model used to describe screening in our graphene/hBN/graphite heterostructure. There are three different mechanisms that contribute to electric polarization responsible for screening. One is due to the intrinsic (interband and intraband) polarizability of graphene bandstructure itself, the other is due to the dielectric permittivity of hBN, the third one is due to image charges on the surface of the graphite gate. Here, we will treat the graphite gate as an ideal conductor, setting electrostatic potential equal zero on the graphite surface. The ideal conductor approximation is justified because the typical screening length in graphite (~1 nm) is much smaller than the hBN spacer thicknesses used in our devices.

We consider the twisted bilayer graphene (TBLG) layer positioned at $z = 0$ and the hBN spacer of width $w$ located beneath it at $-w \leq z \leq 0$. We then expand the potential of a point charge positioned in MAG plane, $\phi(r, z)$, as a sum of Fourier harmonics varying in constant - $z$ planes with coefficients that depend on $z$,

$$\phi(r, z) = \sum_q \phi_q(z) e^{iq \cdot r}.$$

Poisson's equation, written in terms of the quantities $\phi_q(z)$, reads

$$(\partial_z \kappa(z) \partial_z - \kappa(z) q^2) \phi_q(z) = -4\pi (e + \phi_q(z) \Pi_q) \delta(z). \qquad (1)$$

Here, $\kappa(z)$ is the dielectric permittivity of hBN, taken to be $\kappa_{\text{hBN}} \approx 3.5$ for $-w \leq z \leq 0$ and 1 elsewhere, and $\Pi_q$ denotes the intrinsic static polarizability of TBLG.

We solve the 3D Poisson's equation, Eq. 1, for $\phi_q$ vs. $z$, with the ideal conductor boundary condition on the graphite surface, $\phi_q(z = -w) = 0$. From the solution, we determine the potential in the graphene plane:

$$\phi_q(z = 0) = \frac{4\pi e}{q(1 + \kappa_{\text{hBN}} \coth qw) - 4\pi \Pi_q}.$$

We numerically calculate the polarizability of TBLG using the electron bands obtained from the continuum model[1], within the random phase approximation[2]. To suppress the screening effects due to the polarization of the flat bands, we put the Fermi level outside these bands. We then approximate $\Pi_q$ to be isotropic and equal to its value for $q$ in ΓM direction in the Brillouin zone. For very large $q$ ($q \gg \frac{1}{L_M}$, where $L_M$ is the moiré superlattice period), the quantity $\Pi_q$ matches the polarizability value of two electrically decoupled stacked monolayer graphene layers (MLG), $\Pi_q \simeq 2\Pi_{q,\text{MLG}} = -\frac{qe^2}{2\hbar v_F}$, where $v_F \approx 10^6 \frac{m}{s}$ (Fig. S1).

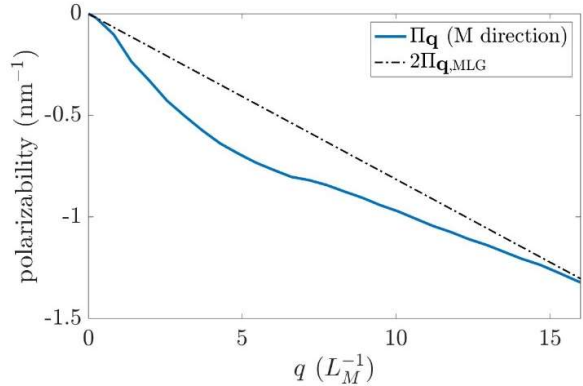

Fig S1. Numerically calculated static polarizability $\Pi_q$ (blue curve) compared with the prediction for two decoupled graphene monolayers at large $q$.

By taking an inverse Fourier transform, the screened interaction of two point charges $e$ at distance $R$ is obtained as

$$V(R) = e^2 \int_0^\infty dq \, \frac{2J_0(qR)}{1+\kappa_{hBN}\coth qw - 4\pi\Pi_q/q}, \qquad (2)$$

Here, $J_n(x)$ is the Bessel function of the first kind. This gives a power-law falloff $V(R) \propto \frac{1}{R}$ for $R \ll w$, and a dependence that decays more rapidly at distances $R > w$ (Fig. S2). There is also a characteristic decrease in the potential in a bulge-shaped feature at distances of the order of $L_M$, arising due to the difference between $\Pi_q$ and $2\Pi_{q,\text{MLG}}$.

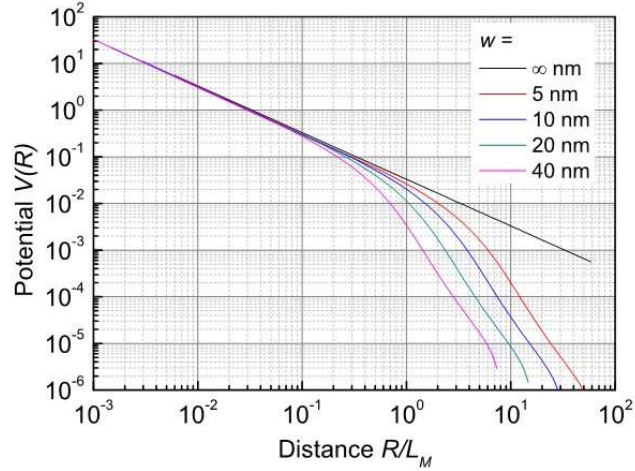

Fig S2. Electron-electron interaction in MAG screened by graphite substrate. Shown is the dependence $V(R)$ vs. $R$ for several different values of the hBN spacer thickness.

Next, we analyze the polarization charge density on graphite surface, $\sigma(r)$, when an electron is localized on a TBLG Wannier orbital, $W(r)$[1]. In order to do so, first, using the solution of Poisson's equation, Eq. 1, and Gauss' law, we can get the polarization charge on graphite surface due to a point charge at $r=0$, $G(r)$,

$$G(r) = -e \int_0^\infty q\,dq \, \frac{\kappa_{hBN} J_0(qr)}{2\pi \sinh qw (1+\kappa_{hBN}\coth qw - 4\pi\,\, q/q)},$$

where $r$ is the lateral distance on the surface of graphite from the point beneath the point charge. Then, we have

$$\sigma(\mathbf{r}) = \int d^2\mathbf{r}' G(|\mathbf{r}-\mathbf{r}'|)|W(\mathbf{r}')|^2.$$

The resulting image charges $\sigma(\mathbf{r})$ are shown in Fig. S12 and Fig. 1A in the main tex.

The on-site Hubbard interaction, $U$, describes the energy cost of adding an electron to one of the Wannier orbitals that already holds one electron. Its dependence on the hBN thickness, $w$, can demonstrate the impact of screening on Mott insulator. Using the potential $V(R)$ in Eq. 2, $U$ can be calculated as

$$U = \int_0^\infty d^2\mathbf{x}\, d^2\mathbf{x}' |W(\mathbf{x})|^2 |W(\mathbf{x}')|^2 V(|\mathbf{x}-\mathbf{x}'|).$$

The Wannier functions have a typical spatial extent of the order of $L_M$. Therefore, when the hBN thickness $w$ is smaller than $L_M$, the screening effects discussed above result in a dramatic suppression of the Hubbard interaction. This behavior, as well as its impact on the phase diagram, is illustrated in Fig.1 of the main text. (Fig. S3)

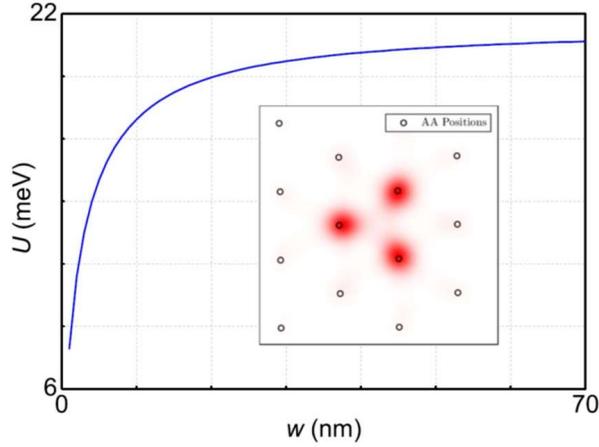

Fig. S3. The Hubbard interaction dependence $U$ vs. hBN thickness $w$, for realistic Wannier functions obtained numerically for the MAG Hamiltonian using the continuum model (shown in the inset). Each Wannier function has three lobes located at AA positions of the moiré superlattice.

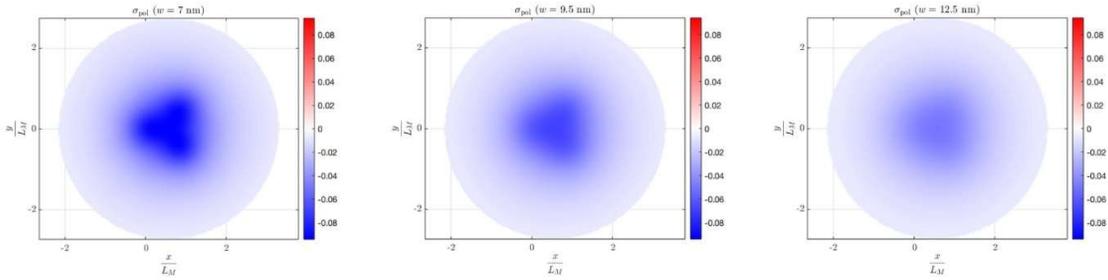

Fig. S4. Polarization charge density calculation for different thicknesses of bottom hBN spacer. $L_M$ is moiré superlattice period, $x$ and $y$ are distances on the two-dimensional plane, $w$ is the thickness of hBN spacer used for calculations. These calculations are used in Fig. 1a in the main text to demonstrate the effect of screening.

### Supplementary Bibliography

1. Koshino, M. *et al.* Maximally Localized Wannier Orbitals and the Extended Hubbard Model for Twisted Bilayer Graphene. *Phys. Rev. X* **8**, (2018).
2. Mahan, G. D. *Many-particle physics / Gerald D. Mahan*. (Plenum Press, 1981).